\documentclass[intlimits,twoside,a4paper]{article}

\usepackage[cp1251]{inputenc}

\usepackage[eqsecnum]{cmpj3}
\usepackage{bm}

\issue{2019}{22}{4}{43606}
\doinumber{10.5488/CMP.22.43606}

\title[X-ray beam induced dynamics in a borate glass]%
{X-rays induced atomic dynamics in a lithium-borate glass}%

\author[F. Dallari \textsl{et al.}]{F. Dallari\refaddr{label1,label2}, G. Pintori\refaddr{label1,label4}, G.Baldi\refaddr{label1}, A. Martinelli\refaddr{label1}, B. Ruta\refaddr{label3,label5}, M. Sprung\refaddr{label2}, G. Monaco\refaddr{label1}
}
\addresses{
\addr{label1} University of Trento, Department of Physics, Via Sommarive, 14, 38123 Povo (TN), Italy
\addr{label2} Deutsches Elektronen-Synchrotron (DESY), Notkestra{\ss}e, 85, 22607 Hamburg, Germany
\addr{label4} University of Trento, Department of Industrial Engineering, Via Sommarive, 9, 38123 Trento, Italy
\addr{label3} ESRF-European Synchrotron Radiation Faculty, 38043 Grenoble, France
\addr{label5} Univ Lyon, Universit\'e Claude Bernard Lyon 1, CNRS, Institut Lumi\`ere Mati\`ere, Villeurbanne, France
}

\usepackage{graphicx}
\usepackage{epstopdf}

\date{Received	July 16, 2019}

\begin{document}

\maketitle

\begin{abstract}
The continuous development of synchrotron-based experimental techniques in the X-ray range provides new possibilities to probe the structure and the dynamics of bulk materials down to inter-atomic distances. However, the interaction of intense X-ray beams with matter can also induce changes in the structure and dynamics of materials. A reversible and non-destructive 
beam induced dynamics has recently been observed in X-ray photon correlation spectroscopy experiments in some oxide glasses at sufficiently low absorbed doses, and is here investigated
in a (Li$_2$O)$_{0.5}$(B$_2$O$_3$)$_{0.5}$ glass. The characteristic time of this induced dynamics is inversely proportional to the intensity of the X-ray beam, with a coefficient that depends on the chemical composition and local structure of the probed glass, making it a potentially new tool to investigate fundamental properties of a large class of disordered systems. While the exact mechanisms behind this phenomenon are yet to be elucidated, we report here on the measurement of the exchanged wave-vector (and thus length-scale) dependence of the characteristic time of this induced dynamics, and show that it follows the same power-law observed in vitreous silica. This supports the idea that a unique explanation for this effect in different oxide glasses should be looked for.

\pacs 61.05.cf, 61.43.Fs, 61.80.Cb
\end{abstract}

\section*{Introduction}

Scattering of coherent electromagnetic radiation allows one to probe spontaneous density fluctuations in the reciprocal space and in the time domain \cite{berne_pecora}.
The experimental techniques that rely on this phenomenon are usually identified with the names of Dynamic Light Scattering (DLS) or Photon Correlation Spectroscopy (PCS) and are nowadays a common tool present in many laboratories for the characterisation of ergodic samples with fast internal dynamics. 
The experimental observable is the normalised correlation function of the scattered intensity $g_2(t)=\langle I_q(0)I_q(t)\rangle/\langle I_q\rangle^2$, where $\langle \dots\rangle$ denotes a temporal average, $q = \frac{4 \piup n}{\lambda} \sin(\theta / 2)$ is the wave-vector exchanged in the scattering process, $\lambda$ is the wavelength of the radiation, $n$ is the refraction index of the material, $\theta$ is the scattering angle and $I_q$ is the intensity scattered at the angle $\theta$ \cite{berne_pecora}.
In normal conditions, $g_2(t)$ is related to the intermediate scattering function, $\Phi_q(t)$, through the Siegert relation $g_2(t)-1=A_q|\Phi_q(t)|^2$, where $A_q\in[0,1]$ is the coherence factor, a set-up dependent parameter.
The extension of the method to the investigation of slow and out of equilibrium systems took advantage of the availability of 2-dimensional detectors such as charge coupled devices (CCD). This has permitted a multispeckle analysis that provides a reliable $g_2$ function for slow dynamical processes in a reduced measurement time by exploiting signal averaging over many independent coherence areas, and has therefore been successfully applied to a variety of undercooled liquids close to the glass transition temperature $T_\text g$  \cite{El_Masri_2009,Dallari2016}. 
Clearly, when studying systems approaching the dynamical arrest, crucial information can be found at length-scales corresponding to the first coordination shells \cite{HANSEN2006178}; therefore, when dealing with molecular or atomic glasses, the limited $q$-range allowed by the use of a visible light represents a strong limitation. The natural way to overcome such a restriction is to reduce the wavelength of the incoming radiation, thus switching to (partially) coherent hard X-rays produced by third (or fourth) generation synchrotron light-sources. In this case, we speak of X-ray Photon Correlation Spectroscopy, or X-PCS, which allows one to probe the atomic dynamics at $q$-values corresponding to inter-atomic distances \cite{Grubel2008}. The price that we have to pay for this extremely wide range of $q$-values is a tremendous drop in the contrast of the detectable $g_2-1$ function as soon as we step outside the small angle X-ray range, i.e., we have to cope with extremely small values of $A_q$ \cite{Abernathy1998}. The only way to mitigate this limitation is to reduce the thickness (along the incident beam direction) of the scattering volume, keeping in mind that thinner samples also means fewer scattered photons and noisier signal. Nevertheless, a compromise can be found, and it is still possible to obtain reasonably good $g_2-1$ functions also at very large $q$'s \cite{Ruta2012}.

X-PCS experiments can be particularly informative to study the glass-transition at the atomic scale. In fact, when a liquid is undercooled at temperatures close to the glass transition temperature, $T_\text g$, its internal dynamics slows down strongly and its characteristic time follows an Arrhenius or even super-Arrhenius laws. This slow-down is expected to take place at all length-scales. It came then as a surprise when, probing the atomic motions at very short length-scales in silicate glasses, Ruta and coworkers measured a relaxation time almost temperature independent and in the range of a few hundred of seconds at temperatures around and below $T_\text g$ \cite{Ruta2014}. In their work a complete decorrelation of the intermediate scattering function was observed even at small fractions of $T_\text g$ indicating an atomic dynamics several orders of magnitude faster than the one observed with macroscopic probes like PCS or viscosity measurements. This unusually fast relaxation was still described by a Kohlrausch-Williams-Watts \cite{KWW} function $\Phi_q(t)=A_q \text{exp}[-(t/\tau)^\beta]$,  with a shape parameter $\beta$ sometimes smaller than 1 \cite{Ruta2014}, corresponding to the usual ``stretched'' shape, and sometimes larger than 1 \cite{Ruta2017}, signalling instead a ``compressed'' shape.
This fast relaxation process was then recognized to be beam-induced \cite{Ruta2017}, with the observed decorrelation time strictly dependent on the intensity of the X-ray beam impinging on the sample. The results found in~\cite{Ruta2017} can be summarised as follows.
a) The characteristic time of this process is inversely proportional to the intensity of the X-ray beam impinging on the sample. 
b) At a fixed intensity, the induced dynamics remains stationary until a certain accumulated dose is reached.
c) The characteristic time of the process depends in a reversible way, and almost ``instantaneously'' on the incident intensity.
d) The shape of the autocorrelation function is independent of the beam intensity, but changes from sample to sample, suggesting that the observed phenomenon is somehow related to the intrinsic dynamics of the glass.
e)~This induced dynamics is observed for network, oxide glasses, while the atomic dynamics of metallic glasses does not appear to be affected by the X-ray beam in X-PCS experiments.

These observations suggest the presence of a complex beam-activated process which differs from the classical radiation damage often encountered in X-PCS studies on soft materials. 
This kind of beam-induced dynamics closely resembles the one observed in transmission electron microscopy experiments performed on thin silica samples \cite{Huang2013}, where the probing particles also act as pump, thus fuelling the observed process. In general, then, this effect may mask or rather go in competition with the spontaneous sample dynamics as reported in \cite{Ruta2012,Pintori2019}. 
However, it has also been observed that the characteristic time of this induced dynamics is inversely proportional to the intensity of the X-ray beam or, more precisely, to the number of X-ray photons absorbed by the sample during one correlation time, with a coefficient that depends on the chemical composition and local structure of the probed glass \cite{Pintori2019,TesiGiovanna}. This observation makes of this effect a potentially new tool to measure fundamental properties of a large group of disordered systems once a clearer understanding of it will be reached. 

In this paper, we report on the observation of beam induced dynamics in a $($Li$_2$O$)$$_{0.5}$$($B$_2$O$_3$$)$$_{0.5}$ glass in X-PCS experiments exploring $q$ values comprised between 1 and 22 nm$^{-1}$. In particular, we discuss here the dose- and $q$-dependence of the characteristic time of this beam induced dynamics, and compare our results with those of previous studies \cite{Ruta2012,Pintori2019}, highlighting the numerous common aspects of this effect that now start to appear. 

\newpage
\subsection*{Experiment}
The data here presented were obtained in two different experiments performed at the beamlines P10 \cite{P10description} and ID10 \cite{ID10description} at the large scale facilities of Petra III (D) and ESRF (F), respectively.
In the experiment at P10, the 8.1 keV monochromatic X-ray beam was focused to a $3\times 3$~\textmu{}m$^2$ spot with an intensity of $\sim 4.5\cdot 10^{10}$ photons/s at 75 mA of current in the storage ring. The detector was a Pixis detector (Princeton Instruments with $1340\times 1300$ pixels, $20\times 20$~\textmu{}m$^2$ pixel size). 
The speckle size, determined by the formula $l_c=1.22\lambda D/d$, where $\lambda=0.15$ nm is the X-ray wavelength, $d$ the X-ray beam diameter at the sample position, and $D=400$ mm the sample-detector distance, was $l_c\sim 24$~\textmu{}m. In this configuration it was possible to cover at the same time a large solid angle and to acquire relatively bright speckle patterns. Thus, the detector area was divided into three equal regions that were analysed independently, corresponding to an angular resolution of $\delta\theta=1.23^\circ$. 
At ID10 the monochromatic, 8.1~keV X-ray beam intensity at the sample position was of $8\cdot 10^{10}$ photons per second at 200 mA of current in the storage ring. The beam spot size at the sample position was of $10\times 8$~\textmu{}m$^2$. The detector was a deep depletion Andor Ikon-M camera. This camera has $1024\times1024$ pixels and a pixel size of 13~\textmu{}m, and was placed at a distance of $D\sim 750$ mm from the sample in order to match  the speckle and  pixel sizes. For the subsequent data analysis, the entire $1024\times1024$ frame was associated to a single $q$, and, therefore, the angular resolution was $\delta\theta\sim1^\circ$.

The samples were prepared according to the following procedure: powders of lithium carbonate and diboron trioxide (both $99\%$ purity, purchased from Sigma-Aldrich) were mixed, finely ground and then melt in a high temperature furnace. The melt was subsequently quenched to room temperature in the shape of small disks (a few mm in diameter) and annealed. The glass disks were then mechanically polished down to $\sim50$~\textmu{}m thicknesses.
During both X-PCS experiments, the samples were kept at a constant temperature of  $300$~K, monitored using a thermocouple inserted close to the sample holder. For both experiments, a photodiode was used to monitor the transmitted beam intensity and to measure the sample thickness, $L$, via the Lambert-Beer law $I=I_0\exp(-\mu L)$, where $I$ and $I_0$ are the transmitted and incident beam intensities, respectively, and $\mu$ is the linear absorption coefficient.

\subsection*{Results and discussion}
\begin{figure}[!t]
\centering
\includegraphics[width=0.9\columnwidth]{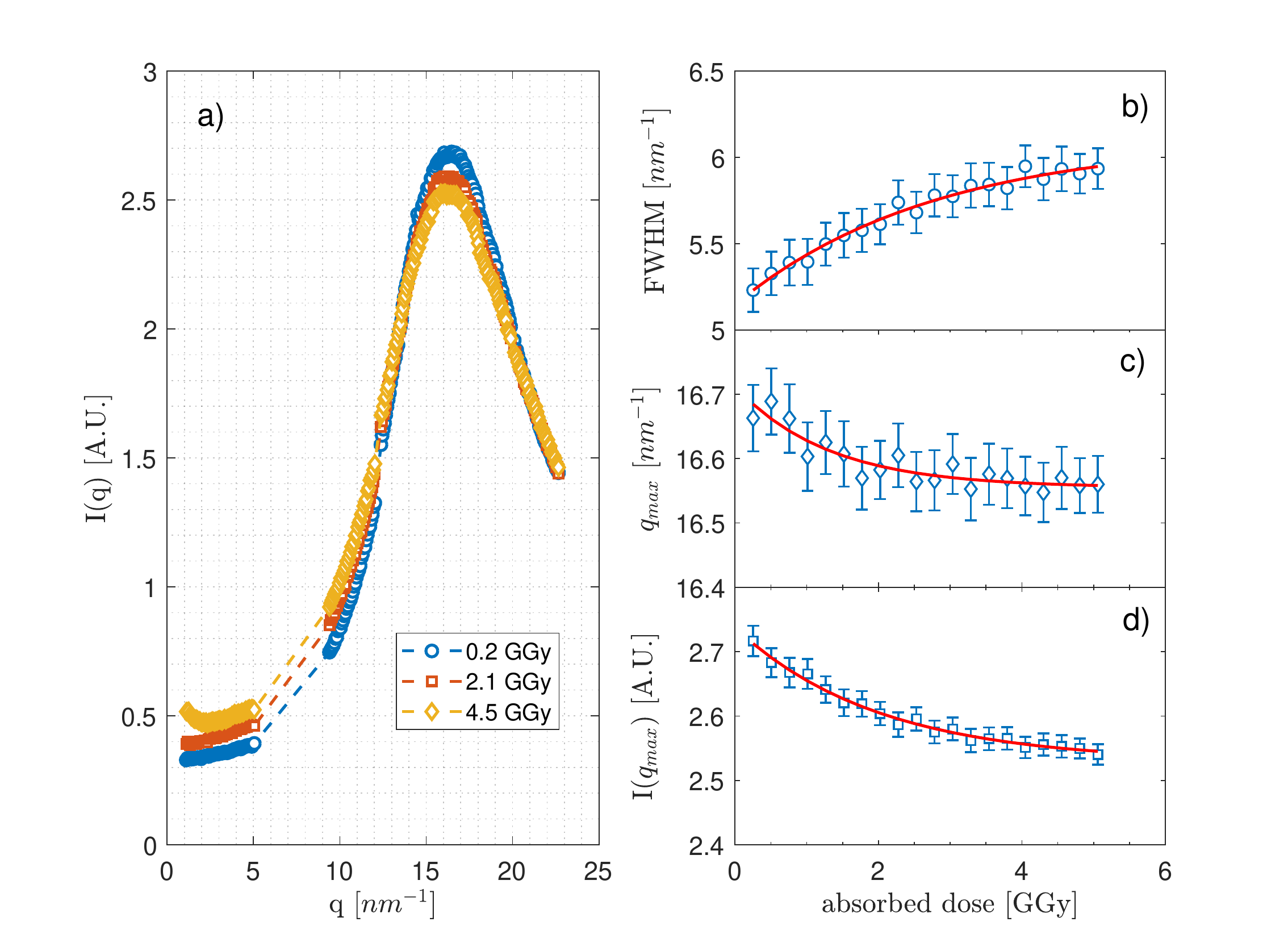}
\caption{\label{fig:sampleDamage}(Colour online) a) Scattered intensity, $I(q)$, at different values of absorbed X-ray radiation immediately after the beginning of the measurement (blue circles, $0.2\cdot10^9~$Gy), at intermediate exposition (red squares, $2.1\cdot10^9$~Gy) and towards the end of the measurement (yellow diamonds, $4.5\cdot10^9$~Gy). The relevant characteristics of the main peak (position, width and height) have been tracked with a Lorentzian fit. On the right, the dose evolution of the main $I(q)$ peak's FWHM (b), position (c) and amplitude (d), are reported. The red lines are exponential fits indicating characteristic doses of $d_\text{FWHM}=(2.8\pm 0.9)$~GGy [corresponding to a characteristic time of $\tau_\text{FWHM}=(1.7\pm 0.5)\cdot10^3$~s for the experimental conditions of the experiment], $d_{q_\text{max}}=(1.3\pm 0.7)$ GGy [corresponding to $\tau_{q_\text{max}}=(0.8\pm 0.4)\cdot10^3$s] and $d_{I(q_\text{max})}=(2.0\pm 0.4)$~GGy [corresponding to $\tau_{I(q_\text{max})}=(1.1\pm 0.2)\cdot10^3$s], respectively. 
}
\end{figure}

Not all oxide glasses have the same resistance to X-rays as the canonical SiO$_2$ glass discussed in \cite{Ruta2017} and some of them, like the $($Li$_2$O$)$$_{0.5}$$($B$_2$O$_3$)$_{0.5}$ glass here studied, show modifications in their local structure already at relatively low absorbed doses as can be seen in figure~\ref{fig:sampleDamage}. There, the scattered intensity, $I(q)$, profiles measured for different values of the total absorbed dose are reported together with a more detailed information concerning the position, amplitude and width of the main peak at different doses. The decrease and broadening of the main peak of $I(q)$ could suggest heating of the scattering volume, but the changes here observed in the peak position are not compatible with previous studies of the temperature dependence of the structure of this lithium borate glass \cite{Majerus2003}. Anyhow, figure~\ref{fig:sampleDamage} indicates a structural modification explainable as an increase in the disorder of the local structure, a consequence of the interaction of the glass with hard X-rays.

\begin{figure}[!t]
\centering
\includegraphics[width=.9\columnwidth]{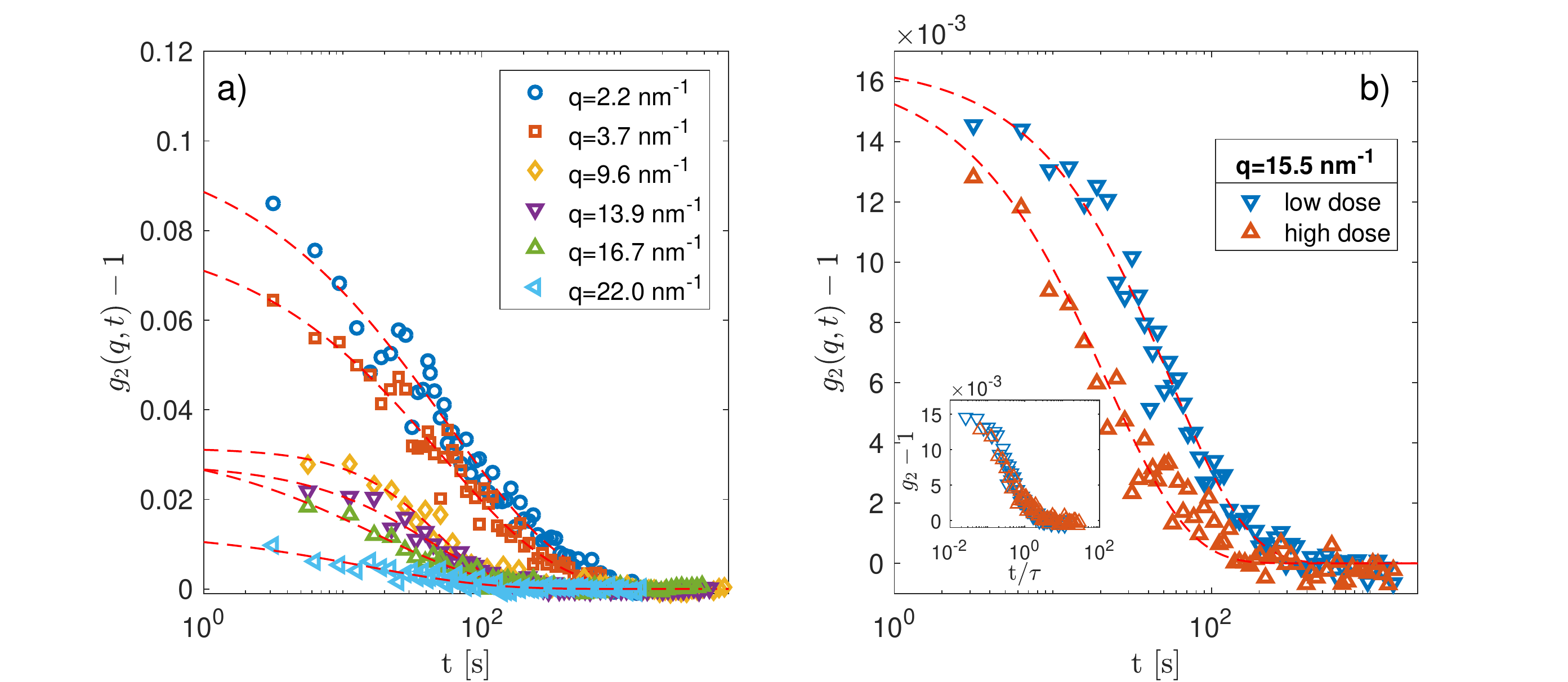}
\caption{\label{fig:g2VariqEldhd}(Colour online) a) Symbols: logarithmically binned autocorrelation functions at different exchanged wavevectors. Dashed lines: stretched exponential fits to the data. The sharp drop in contrast with increasing $q$ is due to the limited longitudinal coherence of the X-ray radiation. These data correspond to a total absorbed dose of $2.1 \cdot10^9$ Gy. b) Correlation function at $q=15.5$~nm$^{-1}$ for two different total absorbed doses. In the inset, the same correlation function is plotted as a function of time rescaled to the corresponding characteristic time. The accumulated dose affects only the timescale of the relaxation process rather than the distribution of the characteristic times $\tau$ as indicated by the fact that once the time scale has been normalised by $\tau$ the two curves overlap.}
\end{figure}

The first set of measurements was performed at P10. Because of the relatively large solid angle collected by the detector, the variation of scattered intensity affects the $g_2-1$ data introducing small changes in the baseline. In order to minimize this effect, each image was normalised with the instantaneous $I(q)$, similarly to what is indicated in \cite{Duri2005}. 
The X-ray's partial coherence leads to a small contrast of the autocorrelation functions as reported in figure~\ref{fig:g2VariqEldhd}~(a). Nevertheless, the collected data are still of rather good quality to perform stretched exponential fits. From these datasets, we can also study the dynamics at different values of the total absorbed dose as illustrated in figure~\ref{fig:g2VariqEldhd}~(b). In order to simplify the discussion, and in view of the overall small dose-related effects on the $g_2-1$ functions, we can compare two different dynamical regimes:  a ``low dose'' (LD) regime associated with a total absorbed dose of $2.1\cdot10^9$ Gy and which in turn corresponds to the regime where the most relevant structural changes in the sample take place; and a ``high dose'' (HD) regime, associated with a total absorbed dose of $4.5\cdot10^9$ Gy and which corresponds to a stationary regime for what concerns the sample's structure.

\begin{figure}[!t]
	\centering
	\includegraphics[width=0.58\columnwidth]{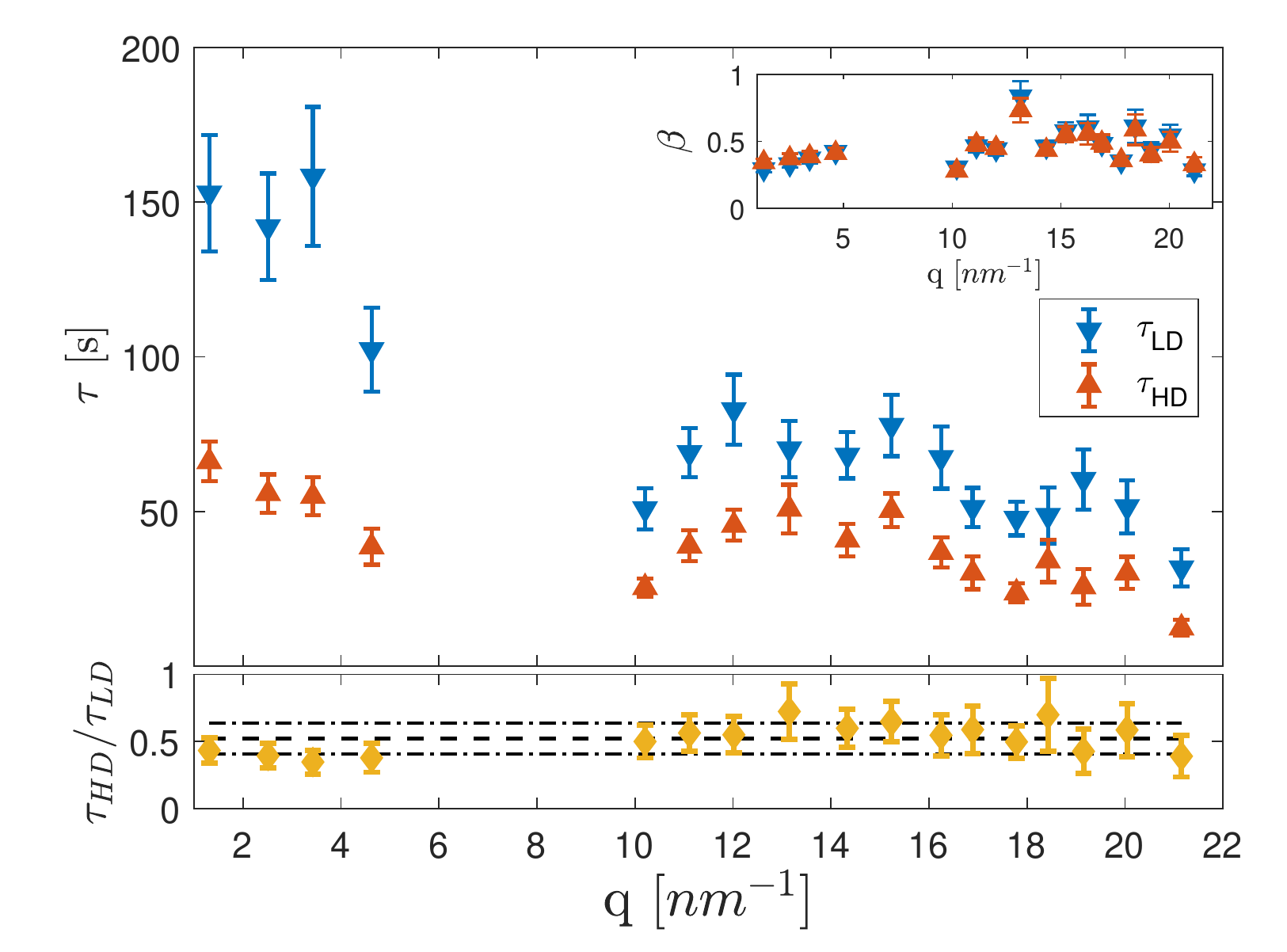}
	\caption{\label{fig:tauVsQ_LiBO1}(Colour online) Top panel: relaxation times obtained for the $($Li$_2$O$)$$_{0.5}$$($B$_2$O$_3$)$_{0.5}$ glass in the two previously defined regimes of accumulated doses: blue down-pointing triangles (low dose) and red up-pointing triangles (high-dose). In the inset, the stretching parameters at different $q$-values for the two different accumulated doses regimes are reported, same symbols as in the main panel. 
		Bottom panel: ratio between the HD and LD relaxation times for all of the investigated $q$ values. The invidual values are consistent within the uncertainty of 1$\sigma$ (dot-dashed line) with the mean value of $\tau_\text{HD}/\tau_\text{LD}=0.5\pm 0.1$ (dashed line). }
\end{figure}

The effect of the irreversible sample damage on the microscopic dynamics is a speed up of the dynamics, while the shape of the correlation curves does not seem to be affected. This is shown in the inset of figure~\ref{fig:g2VariqEldhd}~(b) and, even more explicitly, in figure~\ref{fig:tauVsQ_LiBO1}. Non-exponential relaxations described by KWW functions derive from the average over a distribution of simple relaxation processes \cite{BerberanSantos2005,Patterson1985,Hansen2013}; the invariance of the shape parameter at different absorbed doses then indicates that the permanent sample damage affects in the same way all relaxation processes taking place in the scattering volume, maintaining untouched their distribution. 
In figure~\ref{fig:tauVsQ_LiBO1}, the observed relaxation times and stretching exponents for LD and HD conditions are reported. These two sets of relaxation times appear to maintain the same general features as a function of $q$: in both cases the dynamics is described by a stretched exponential, with the characteristic time showing a slight slow down in correspondence of the main $I(q)$ peak, reminiscent of a de Gennes narrowing effect \cite{DEGENNES1959825} that one would expect in the frequency domain, on top of a continuous speed up trend on increasing $q$. 
The ratio between the relaxation times related to the LD and HD regimes appear to be substantially constant vs $q$, as shown in the bottom panel of figure~\ref{fig:tauVsQ_LiBO1}, suggesting that the effects of the irreversible structural modification (sample damage) are length-scale invariant in the probed dose- and $q$-range. Lastly, the fact that the dynamics speeds up in an increasingly disordered sample confirms the observations reported in \cite{TesiGiovanna,Pintori2019}, in which a strong connection between the time-scale of the beam induced dynamics and the local structure of the glass has been observed.

It can now be useful to clarify the physical quantities probed in this kind of experiment when a beam-induced dynamics is detected \cite{Pintori2019}. When we perform a photon-correlation experiment, what is measured is the time, $\tau$, required by the system to completely rearrange its internal structure. Thus, if the process can be described by a simple relaxation process, we can say that, after a time $\tau$, only a fraction of particles in the scattering volume equal to $1/\re$ has not moved over the length-scale defined by $1/q$. 
From \cite{Pintori2019}, we know that the characteristic time corresponding to beam-induced dynamics is a function of the total absorbed X-ray radiation and can be expressed by the equation:
\begin{equation}\label{eq:Nunits}
\frac{\tau}{\tau_\text{lag}}= \frac{1}{\re}\frac{U/ N_\text{units}}{\tau_\text{exp}(1-\re^{-L\mu})I_0}\,,
\end{equation}
where $\tau_\text{lag}$ is the elapsed time between two subsequent images, $\tau_\text{exp}$ is the exposure time for a single image and $U$ is the number of elementary units (e.g., atoms) present in the scattering volume. A good evaluation of this quantity requires the precise knowledge of parameters like the beam size and the sample thickness. Lastly, $N_\text{units}$ is the number of elementary units (e.g., atoms) that move by $1/q$ after the absorption of one photon. This latter quantity is a property specific of the sample that can be determined only with X-PCS experiments: it is actually the real outcome of an X-PCS experiment in beam-induced conditions. Moreover, given its very definition, it is the ideal observable to be used when comparing results obtained in different experimental conditions (i.e., on different beamlines). 

The meaning of $N_\text{units}$ is still under study. One of the possible explanations is that the absorbed X-ray photon generates a highly energetic photo-electron that travels in the glass network interacting inelastically with a number of atoms and, therefore, inducing small displacements along its path. In this framework, $N_\text{units}$ can be seen as an indirect measurement of the photo-electron life-time and the photon energy will result distributed along a relatively large volume of the sample. In an alternative interpretation~\cite{Pintori2019}, the energy deposited by the X-ray photon triggers the relaxation of local stresses stored inside the glass after its quench. In this framework, $N_\text{units}$ would then be a measurement of a relevant sample property depending also on its thermal history, and the deposited energy would remain confined in a narrow volume. With the currently available information, it is still not possible to clearly discriminate between these two interpretations, and a systematic study of the $N_\text{units}$ dependence on parameters such as chemical composition, local structure, thermal history and temperature is required \cite{TesiGiovanna,Pintori2019}.

Because of the constraints due to the vacuum chamber used for the experiment at P10, it was not possible to acquire data between 5 and 10 nm$^{-1}$. A follow-up experiment at ID10 was then performed to cover the missing $q$ range.

As explained above, the beam induced dynamics crucially depends on a number of experimental parameters and the best way to compare different datasets that come from different instruments is to use $N_\text{units}$ as defined in equation~\ref{eq:Nunits}. For materials with relatively small absorption coefficients (like lithium borate glasses) the precision required for the measurement of the sample thickness $L$ is not crucial, as can easily be demonstrated with a series expansion of the exponential at the denominator in equation~\ref{eq:Nunits}. On the other hand, experimental parameters such as the total intensity of the X-ray beam impinging on the sample and the spot size of the focused X-ray beam, play a crucial role in the determination of $N_\text{units}$, and might carry some uncertainties that would result in significant systematic errors. To this purpose, a number of measurements on the same silica sample at the same exchanged wavevector have been performed at the two different beamlines. The values obtained in different experiments with the same instrument have been averaged, and the ratio between the two values of $N_\text{units}$ has been found to be $N_\text{units}^\text{ESRF}/N_\text{units}^\text{Petra}=1.3\pm0.3$, thus compatible with 1 within 1$\sigma$. This result at the same time validates the model of equation~\ref{eq:Nunits} and the measurements of the absolute intensity available at both beamlines, and gives us confidence to treat datasets obtained at ID10 and P10 in a combined way.

\begin{figure}[!t]
	\centering
	\includegraphics[width=0.75\columnwidth]{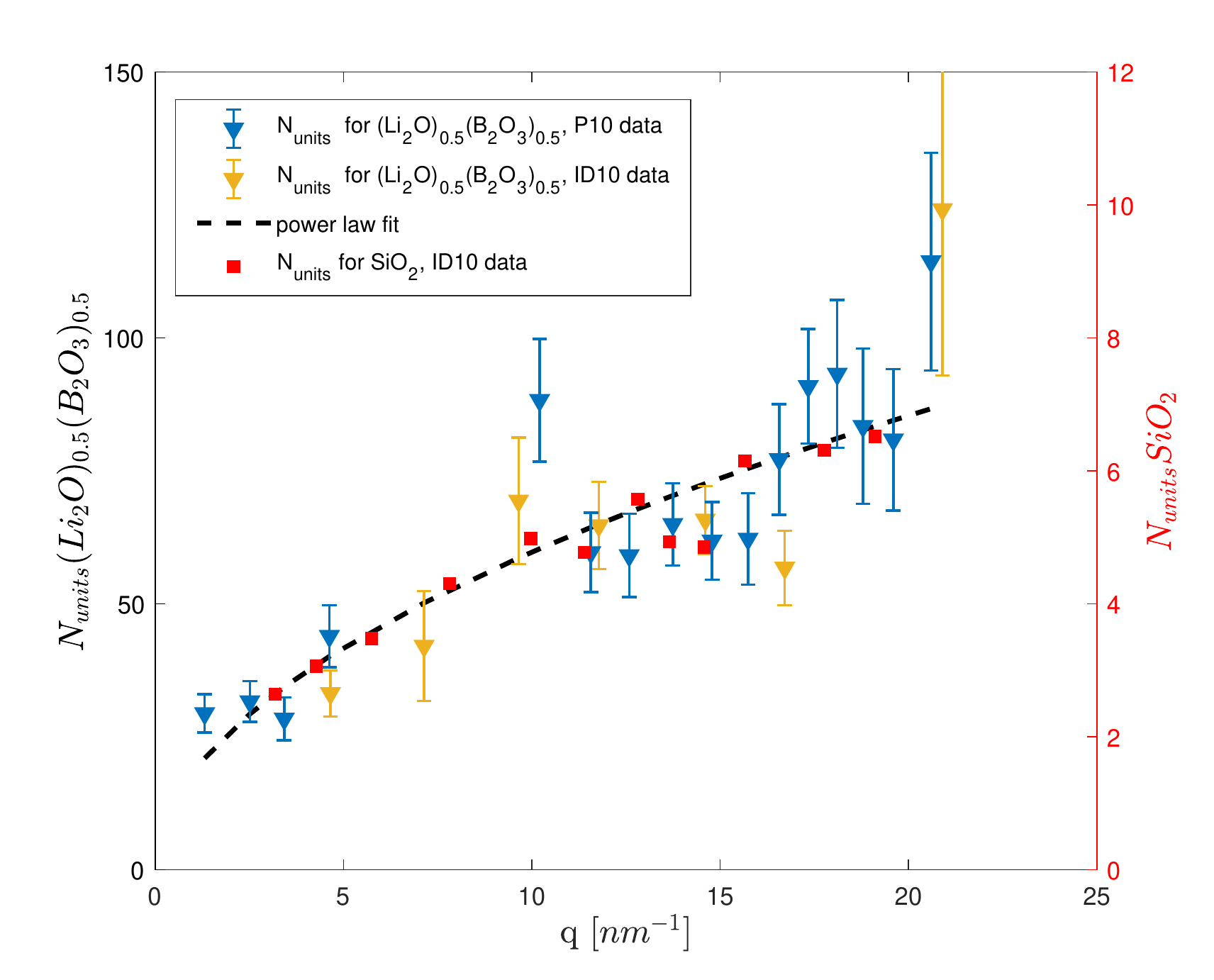}
	\caption{\label{fig:confrontoNunitsVsQ_petraESRFLiBO}(Colour online) Comparison of the number of $($Li$_2$O$)$$_{0.5}$$($B$_2$O$_3$)$_{0.5}$ units, $N_\text{units}$, that move by $1/q$ after the absorption of a single 8.1 keV X-ray photon for a $($Li$_2$O$)$$_{0.5}$$($B$_2$O$_3$)$_{0.5}$ glass obtained in two separate experimental runs at two different beamlines (P10 at Petra III and ID10 at the ESRF). Blue triangles: $N_\text{units}$ data obtained from the LD relaxation times reported in figure~\ref{fig:tauVsQ_LiBO1}; dark-yellow triangles: data obtained from a set of measurements carried out at ID10 on the same kind of lithium borate glass. Dashed
		line: power law fit ($N_\text{units}=N_0\cdot q^\alpha$) to both low-dose datasets, with $N_0=18\pm9$ and $\alpha=0.5\pm0.2$. Red squares (right-hand axis): SiO$_2$ $N_\text{units}$ relative to the SiO$_2$ glass extracted from the beam-induced relaxation times reported in \cite{Ruta2017}. The values of $N_\text{units}$ relative to silica are nearly a factor 10
		smaller than those corresponding to the lithium borate. This behavior is expected as the local structures of these glasses are different, and is in agreement with the results obtained from the experiment performed on borate glasses with different local structures reported in \cite{TesiGiovanna}. 
	}
\end{figure}

For the second run of measurements at ID10, a new sample of the $($Li$_2$O$)$$_{0.5}$$($B$_2$O$_3$)$_{0.5}$ glass was prepared closely following the same procedure as for the P10 run and applying the same thermal treatments. Then, knowing how to compare the data from the two different beamlines, we could merge the two datasets of $N_\text{units}$ values. The result, reported in figure~\ref{fig:confrontoNunitsVsQ_petraESRFLiBO}, shows a reasonably good agreement between the low dose data obtained at Petra III (blue triangles) and the ones that came from ESRF (dark yellow circles). The combination of these data can be described in first approximation by a power law $N_\text{units}=N_0\cdot q^\alpha$ with exponent $\alpha=0.5\pm0.2$. The de Gennes narrowing effect on $\tau$ commented in figure~\ref{fig:tauVsQ_LiBO1} translates in terms of $N_\text{units}$ into a plateau region that starts at $\sim 10$ nm$^{-1}$. This tendency to reach an upper value in $N_\text{units}$ can be explained taking into account the fact that at high $q$-values we are probing distances smaller than the characteristic X-ray induced atomic movement, in accordance with the predictions of a phenomenological model developed for colloidal gels \cite{Trappe2007}.
Interestingly, a power law with the same exponent can be used to describe the $q$-dependence of the $N_\text{units}$ parameter extracted from the data available in \cite{Ruta2017} for the case of the silica glass. As shown in figure~\ref{fig:confrontoNunitsVsQ_petraESRFLiBO}, in fact, the $q$-dependence of the number of units that move after the absorption of one X-ray photon for the lithium borate glass and for silica scale one on top of the other, i.e., they differ for a $q$-independent multiplication factor.  
There is, however, an interesting difference between the induced dynamics in silica and in the lithium borate glass here reported since in the former case the correlation functions are compressed, with $\beta \sim 1.5$, while in the latter case we observe a stretched exponential decay (see the inset of figure~\ref{fig:tauVsQ_LiBO1}).

\newpage
\section*{Conclusions}
In conclusion, we have demonstrated that, even though the beam induced dynamics effect is strongly set-up dependent, we can normalise our data in order to extract a material-dependent quantity, $N_\text{units}$, which corresponds to the number of units that move after the absorption of a single X-ray photon. Moreover, we have seen that different samples produced following the same thermal protocol respond in the same way to the X-rays. We have also seen that, for $($Li$_2$O$)$$_{0.5}$$($B$_2$O$_3$)$_{0.5}$, the irreversible sample damage occurs rather quickly compared to other network glasses and that this permanent modification affects the reversible beam induced dynamics with a speed-up of the characteristic decay time at all the the probed length-scales. Moreover, the irreversible sample damage does not alter the distribution of the individual relaxation processes. 

The number of units that move following an absorption event appears to be weakly $q$-dependent in the $q$-range probed in the present work, with $N_\text{units}\sim q^{0.5}$. The same $q$ dependence can be also found for $N_\text{units}^{\text{SiO}_2}$ derived from \cite{Ruta2017}, suggesting that the general characteristics of the beam induced dynamics are the same in different materials. One additional interesting piece of information comes from the different shape of the correlation curves between the mechanically weaker lithium borate glass and the more rigid vitreous silica. 
 
In the stress-relaxation interpretation of $N_\text{units}$, the beam-induced dynamics might be represented, at least qualitatively, with the same models developed for soft colloids and gels \cite{Bouzid2017}. In recent simulations, it has been shown that the shape parameter, as well as the $q$-dependence of $\tau$, depends on the ratio between the energy of the stress heterogeneity frozen-in during the solidification of the glass and the thermal energy, that could be associated, in the experiments discussed here, with the energy deposited by the X-ray photons \cite{Bouzid2017}. For rather small values of this ratio, stretched intermediate scattering functions can be observed \cite{Bouzid2017}, suggesting that the relative amount of energy accumulated as microscopic stresses in the lithium borate glasses was smaller than that in the vitreous silica sample measured in \cite{Ruta2017}. This intriguing interpretation requires, however, to be tested against more experimental data.

\section*{Acknowledgements}
Parts of this research were carried out at the beamline P10 in the experiment number: I-20150305 EC, at DESY, a member of the Helmholtz Association (HGF).

The research leading to this result has been supported by the project CALIPSOplus under the Grant Agreement 730872 from the EU Framework Programme for Research and Innovation HORIZON 2020.

Parts of this research were also performed at beamline ID10 during the experiment
SC-4541 at the European Synchrotron Radiation Facility (ESRF), Grenoble, France. We are grateful to Federico Zontone at the ESRF for providing assistance in using beamline ID10.

\ukrainianpart 

\title{Індукована рентгенівськими променями атомна динаміка в літій-боратному склі} 

\author{Ф. Далларі\refaddr{label1,label2}, Д. Пінторі\refaddr{label1,label3}, Д. Бальді\refaddr{label1}, А. Мартінеллі\refaddr{label1},
	Б. Рута\refaddr{label4,label5}, М. Шпрунг\refaddr{label2}, Д.~Монако\refaddr{label1}}

\addresses{
	\addr{label1} Університет Тренто, Фізичний факультет, 38123 Пово, Італія
	\addr{label2} Німецький Електрони-Синхротрон (DESY), 22607 Гамбург, Німеччина
	\addr{label3} Університет Тренто, Факультет промислової інженерії, 38123 Тренто, Італія
	\addr{label4} ESRF-Європейський центр синхротронної радіації, 38043 Гренобль, Франція
	\addr{label5} Університет Ліону 1 ім. Клода Бернара, CNRS, Інститут світла і матерії, Вільюрбан, Франція
}

\makeukrtitle

\begin{abstract}
	Постійний розвиток експериментальних технік на основі синхротрона в рентгенівській області забезпечує
	нові можливості дослідити структуру та динаміку об'ємних матеріалів аж до міжатомарних відстаней.
	Однак, взаємодія інтенсивних пучків рентгенівських променів з матерією може також індукувати зміни в 
	структурі та динаміці матеріалів. Зворотня та індукована недеструктивним пучком динаміка спостерігалась
	недавно в експериментах по рентгенівській фотонній кореляційній спектроскопії в деяких оксидних скловидних
	системах при достатньо низьких абсорбованих дозах, та досліджено тут у склі $($Li$_2$O$)$$_{0.5}$$($B$_2$O$_3$)$_{0.5}$. 
	Характеристичний час цієї індукованої динаміки є зворотньо пропорційний до інтенсивності рентгенівського 
	пучка, з коефіцієнтом, який залежить від хімічного складу та локальної структури досліджуваного скла, роблячи
	її потенціально новим засобом для дослідження фундаментальних властивостей великого класу невпорядкованих 
	систем. Хоча точні механізми цього явища ще потрібно встановити, ми повідомляємо тут про вимірювання 
	залежності характеристичного часу цієї індукованої динаміки від хвильового вектору передачі (та відповідного
	масштабу довжини), і показуємо, що вона слідує тому ж степеневому закону, що спостерігається для аморфного 
	кремнезему. Це підтверджує ідею, що потрібно шукати однакове пояснення цього ефекту в різних оксидних 
	скловидних системах.
	
	\keywords  cкловидні системи, перехід в стан скла, рентгенівська фотонна кореляційна спектроскопія, когерентне розсіювання рентгенівських променів 
	
\end{abstract}

\end{document}